\documentclass[%
 reprint,
superscriptaddress,
 amsmath,amssymb,
 aps,
]{revtex4-1}

\usepackage{booktabs}
\usepackage{multirow}

\usepackage{subcaption}

\usepackage{bm} 
\usepackage{graphicx}
\usepackage{dcolumn}
\usepackage{bm}
\usepackage{hyperref}

\begin{document}

\preprint{APS/123-QED}

\title{Exploring the Possibility of Testing the No-Hair Theorem with Minkowski-deformed Regular Hairy Black Holes Via Photon Rings}

\author{Xilong Yang}

\author{Meirong Tang}

\author{Zhaoyi Xu}%
\email{zyxu@gzu.edu.cn(Corresponding author)}
\affiliation{%
 College of Physics,Guizhou University,Guiyang,550025,China
}%


\begin{abstract}
In this paper, we investigate the optical appearance of a regular static spherically symmetric hairy black hole within the context of Minkowski deformation governed by the parameter \(\alpha\). This black hole describes a hairy black hole with geometric deformations in the radial and temporal metric components, parameterized by \(\alpha\). The optical appearance of the black hole, illuminated by a static thin accretion disk in three toy emission function models, exhibits distinctive shadows and photon rings. Our findings reveal that for static spherically symmetric hairy Minkowski-deformed regular black holes, the event horizon radius \(r_h\), photon sphere radius \(r_{ph}\), critical impact parameter \(b_{ph}\), and innermost stable circular orbit radius \(r_{isco}\) all have a positive correlation with \(\alpha\). These parameters affect the null geodesic trajectories, shadows, and the optical appearance of the photon rings illuminated by a static thin accretion disk around the black hole. Utilizing this, we obtained the optical appearance of this black hole when observed in the forward direction of the optical and geometric thin accretion disk models. In all three models, the radius of the photon ring images also shows a positive correlation with the parameter \(\alpha\). Therefore, under the static spherically symmetric hairy Minkowski-deformed regular black hole, there is no degeneracy of photon rings dependent on the parameter \(\alpha\). Theoretically, this allows for the distinction of different spacetime metrics of hairy Minkowski-deformed regular black holes, providing a potential method for testing the no-hair theorem through future observations of black hole photon rings.

\begin{description}
\item[Keywords]
No-hair Theorem, Shadows, Photon Rings, Regular Hairy Black Holes, Null Geodesics.
\end{description}
\end{abstract}

\maketitle


\section{\label{sec:level1}Introduction}
In modern theoretical physics, black holes, as one of the most mysterious celestial bodies in the universe, have always been a focal point of research due to their intriguing properties and structures. The optical appearance of black holes provides an important window into understanding these enigmatic entities. In 2019, the Event Horizon Telescope (EHT) collaboration achieved the first detection of the accretion flow around the supermassive object at the center of the $M87^*$ galaxy and released the "image" of the $M87^*$ black hole \cite{event2019first, akiyama2019first}, marking a significant starting point for testing the no-hair theorem \cite{gan2021photon}. Subsequently, in 2022, the EHT collaboration released the image of $Sgr A^*$, the supermassive black hole at the center of our galaxy \cite{grigorian2024relationship, lu2018detection, akiyama2022first, event2023first}. The primary goal of the EHT is to observe supermassive black holes at the centers of galaxies. It achieves this by combining data from eight radio telescopes distributed globally, creating a virtual telescope with an effective aperture equivalent to the diameter of the Earth, thus enhancing the angular resolution of the telescope to a level sufficient to observe structures at the scale of the event horizon\cite{akiyama2019first, event2019first}.

The photon ring, formed by photons orbiting the black hole, is a critical component of the black hole shadow. It offers a direct method to observe and study the strong gravitational field effects of black holes 	\cite{gralla2019black} Research on the photon ring of black holes is a frontier topic in theoretical and astrophysics, attracting wide attention\cite{narayan2019shadow, johannsen2013photon, johannsen2013photon}. With advances in observational technology, particularly the success of the EHT project, the understanding of photon rings around black holes has reached unprecedented depths\cite{event2019first}. In recent years, through observations of photon rings, scientists have gained deeper insights into the structure of black hole accretion disks, gravitational lensing effects, and the rotational speed of black holes\cite{sui2024effect, gan2021photon, meng2024images, yang2023shadow}. The study of higher-order ring images in black hole accretion disks also makes it possible to distinguish different accretion models\cite{bisnovatyi2022analytical}. These studies not only deepen the understanding of the predictions of general relativity but also provide crucial clues for testing the no-hair theorem and exploring quantum gravity theories. Furthermore, the study of photon rings is vital for understanding the physical processes under extreme conditions in the universe, promoting in-depth exploration of high-energy astrophysics, cosmological principles, and fundamental physical laws\cite{gussmann2021polarimetric}. In the future, with technological advancements and the deployment of new astronomical observation instruments, research on photon rings will continue to expand our understanding of the profound mysteries of the universe.

The optical properties of supermassive objects were first systematically analyzed by Bardeen, who examined the shadow and photon rings of black holes\cite{bardeen1973timelike}. Building on this foundation, Gralla et al. explored the shadow and photon rings of a Schwarzschild black hole within the context of an optically thin accretion disk model, revealing that both the position and brightness of the photon rings exhibit logarithmic divergence.\cite{gralla2019black}. Merce Guerrero et al. used ray-tracing methods to study the shadow and optical appearance of a black bounce illuminated by a thin accretion disk\cite{guerrero2021shadows}. Abdujabbarov et al. investigated the shadow of a charged rotating black hole, finding that, besides the black hole's angular momentum, the gravitational electric charge term also alters the shape of the black hole shadow\cite{abdujabbarov2013shadow}. Saleem and Aslam studied the optical properties of a charged non-commutative Kiselev black hole under different accretion material configurations, discovering that the size of the black hole shadow and its related attributes are closely related to the black hole's state parameters and spacetime geometry\cite{saleem2023observable}. Takahashi analyzed the shadow of a rotating black hole in an accretion disk, finding that it is challenging to determine the black hole's spin parameter from the size and shape of the black hole shadow in the accretion disk\cite{takahashi2004shapes}. Sergio V. M. C. B. Xavier et al. studied the photon rings of traversable wormholes\cite{xavier2024traversable}.

The no-hair theorem is a fundamental theorem in relativistic black hole theory, stating that the only external attributes of a black hole are those related to the properties of massless fields allowed by conserved current integrals, i.e., the physical properties of a black hole are determined solely by three parameters: mass, charge, and angular momentum. This implies that all other information about a black hole is lost during its formation\cite{adler1978no}. However, the validity and universality of this theorem remain open questions in theoretical physics. On one hand, in strong gravitational fields, it seems possible to identify black hole shadows illuminated by accretion disks through photon rings. For instance, in the progress of no-hair theorem research, Tim Johannsen studied the effects of two deviations on the photon rings of Kerr and quasi-Kerr black holes, which would lead to asymmetry in the ring shape\cite{johannsen2013photon}. Övgün et al. investigated the shadow and energy emission rate of spherically symmetric black holes under non-commutative geometry within the Rastall gravity framework, providing valuable information about compact objects like black holes\cite{ovgun2020shadow}. On the other hand, Maxim Lyutikov and Jonathan C. found that the no-hair theorem does not formally apply to black holes formed by the collapse of rotating neutron stars\cite{lyutikov2011slowly}. Juan Calderón Bustillo et al. found that applying quasi-normal modes to GW150914 could not provide strong evidence supporting the no-hair theorem\cite{bustillo2020black}.With the discovery of various "hairy" black hole solutions—those requiring additional parameters not related to conserved charges to complete the required black hole solution—a significant window has been opened for studying potential violations of the no-hair theorem. These solutions include, for example, "colored black holes," which are solutions to the static, spherically symmetric Einstein-Yang-Mills-Higgs equations with an SU(2) gauge group\cite{bizon1990colored}; black hole solutions in spontaneously broken classical Einstein-Yang-Mills-Higgs theory\cite{greene1993eluding}\cite{mavromatos1996aspects}; black hole solutions with "Skyrme hair" in nonlinear matter fields\cite{droz1991new}; and non-Abelian black holes in the Einstein-Yang-Mills theory\cite{volkov1999gravitating}, among others.

In this context, the black hole model studied in this paper is the regular hairy black hole model with Minkowski deformation proposed by Jorge Ovalle et al\cite{ovalle2023regular}. Darío Núñez et al. found that the non-trivial structure region of nonlinear matter fields must extend beyond 3/2 times the horizon radius, independent of all other parameters present in the theory\cite{N_ez_1996}. Building on this, Pabitra Tripathy discovered that, regardless of the dimension or Lovelock order, the hair of a static spherically symmetric black hole must extend at least to the photon sphere\cite{tripathy2024lower}. Therefore, this paper aims to explore the feasibility of testing the no-hair theorem by observing and analyzing the behavior and characteristics of photon rings. This method relies on the precise calculation of null geodesics in strong gravitational fields and the radiation behavior of the accretion disk. Through these studies, we aim to better understand the properties of black holes, verify the universality of the no-hair theorem, and further test the theories of general relativity and quantum gravity. The advancements in this research field are not only significant for the development of theoretical physics but also have broad impacts on astrophysics, cosmology, and high-energy physics experiments. By studying photon rings and Minkowski-deformed hairy black holes, we can gain a deeper understanding of the properties of these mysterious celestial objects in the universe and potentially pave the way for the discovery of new physical theories.

The structure of this paper is as follows: Section \ref{sec:level2} introduces the regular hairy black holes with Minkowski deformation, discussing their event horizon radius and effective potential under the influence of parameter \(\alpha\) and the corresponding minimum unstable orbit radius. In Section \ref{sec:level3}, we derive the corresponding null geodesic equations and, assuming an optically thin accretion disk around the black hole, provide the impact parameter curves influenced by parameter \(\alpha\) for different null geodesics and the corresponding geodesic trajectories. We also re-constrain the parameter \(\alpha\) based on the shadow structure of M87* and Sagittarius A* black holes observed by the EHT, making it more practically significant. Section \ref{sec:level4} presents the shadow, photon ring, and corresponding observed intensity flux diagrams for distant observers. Finally, Section \ref{sec:level5} concludes with a summary and discussion. Throughout this paper, we adopt the geometric unit system with \( c = G = M = 1 \). The metric signature used is \((- , +, +, +)\).

\section{\label{sec:level2}Minkowski-Deformed Regular Hairy Black Hole}
In this section, we first review the spherically symmetric regular hairy black hole obtained through Minkowski deformation. By employing the gravitational decoupling (GD) method, a hairy black hole without curvature singularity has been proposed in \cite{ovalle2023regular}. This study identifies the matter source that induces the Minkowski vacuum deformation, with the maximum static deformation corresponding to the Schwarzschild solution and the maximum steady-state deformation corresponding to the Kerr metric. This paper focuses on the static case. In the Schwarzschild-like coordinates, the line element of this regular black hole can be written as
\begin{equation}
ds^2 = -e^{\nu(r)} dt^2 + e^{\lambda(r)} dr^2 + r^2 d\Omega^2,
\label{1}
\end{equation}
where \(d\Omega^2 = d\theta^2 + \sin^2\theta d\phi^2\), \(\nu(r)\) and \(\lambda(r)\) are functions of the radial coordinate \(r\). For a spherically symmetric case, we have
\begin{equation}
e^{\nu} = e^{-\lambda} = 1 - \frac{2M}{r} + \frac{e^{-\alpha r / M}}{rM} (\alpha^2 r^2 + 2M\alpha r + 2M^2).
\label{2}
\end{equation}

From Eq. \ref{2}, it can be seen that the metric differs from the Schwarzschild metric only by the introduction of the parameter $\alpha$, which directly represents the hairy deformation of the black hole. It signifies the geometric deformation in the radial and temporal metric components. When $\alpha \rightarrow 0$, the Minkowski spacetime is recovered; likewise, when $\alpha \rightarrow +\infty$, the Schwarzschild solution is restored. Thus, the possible horizon for this metric is obtained by finding the solution $r_h = r_h(\alpha)$ to

\begin{equation}
e^{-\lambda(r_h)} = 0. 
\label{3}
\end{equation}

\begin{figure}[ht]
\centering
\includegraphics[width=0.5\textwidth]{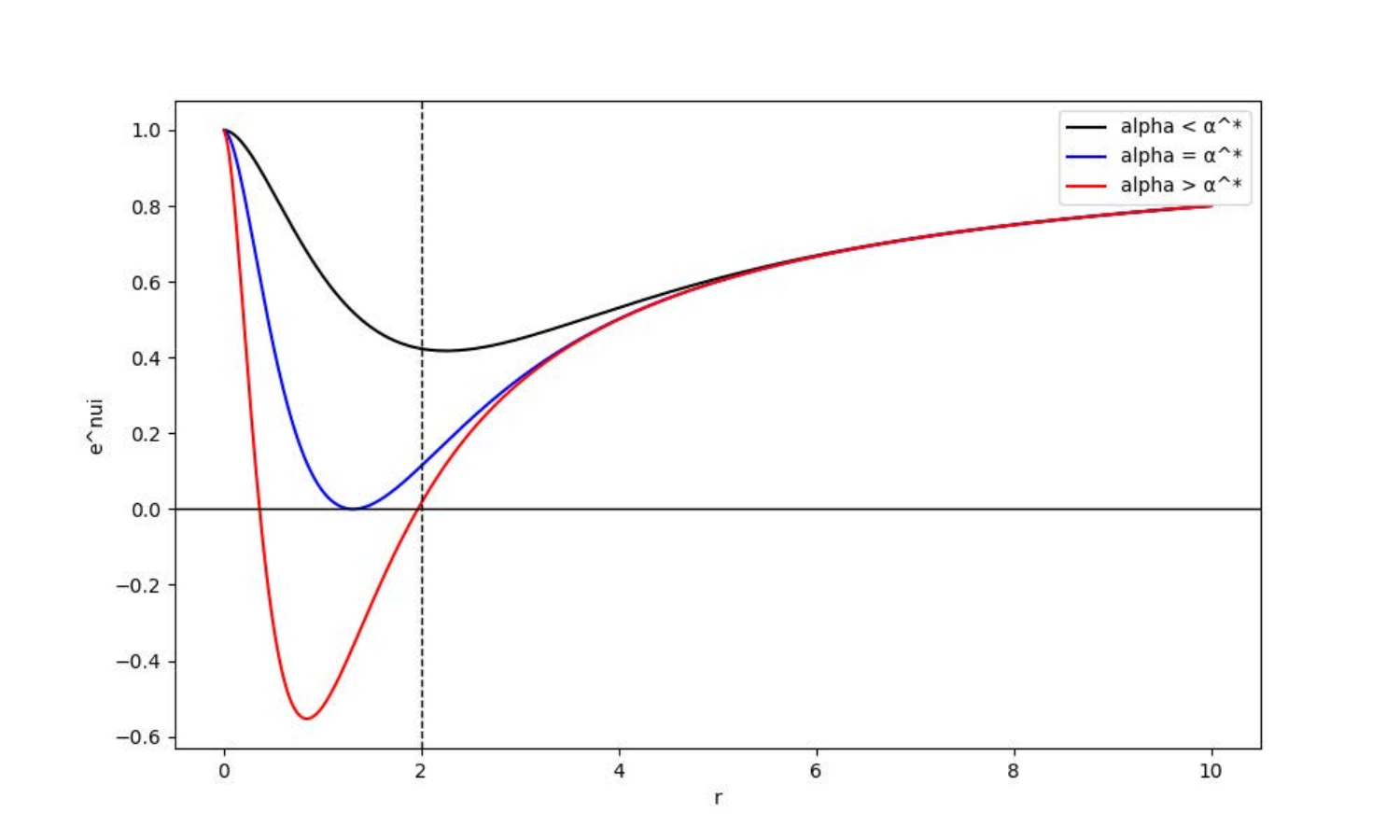}
\caption{The effective potential $V_{\text{eff}}(r)$ versus $r$ for the Schwarzschild black hole, showing the relationship between $r_{\text{ph}}$ and $b_{\text{ph}}$. The event horizon radius is $r_h = 2$ and the photon sphere radius is $r_{\text{ph}} = 3$.}
\label{fig:1}
\end{figure}

The analysis of Eq. \ref{3} reveals three scenarios: a critical value $\alpha = \alpha^*$; a case where $\alpha < \alpha^*$ has no zero points; and a case where $\alpha > \alpha^*$ has two zero points, which are the Cauchy horizon and the event horizon, respectively, as shown in Figure \ref{fig:1}. These three scenarios are displayed as follows: no event horizon, the critical case, and a black hole with both a Cauchy horizon and an event horizon. For the critical value, $\alpha^* \approx 2.575$.

According to the cosmic censorship hypothesis, which prohibits the formation of naked singularities\cite{PhysRevLett.14.57}, if a black hole horizon exists, the range of the parameter $\alpha$ is from $\alpha^*$ to positive infinity.

\section{\label{sec:level3}Geodesic Equations and Effective Potential}

To study the behavior of photons around the Minkowski-deformed regular hairy black hole, it is necessary to calculate the geodesic motion of the photons. In the case of a spherically symmetric spacetime, we first need to describe the Lagrangian to obtain the constants of motion (the energy and angular momentum of the photon along the symmetry axis) in relation to the generalized momentum.

\begin{figure}[ht]
\centering
\includegraphics[width=0.5\textwidth]{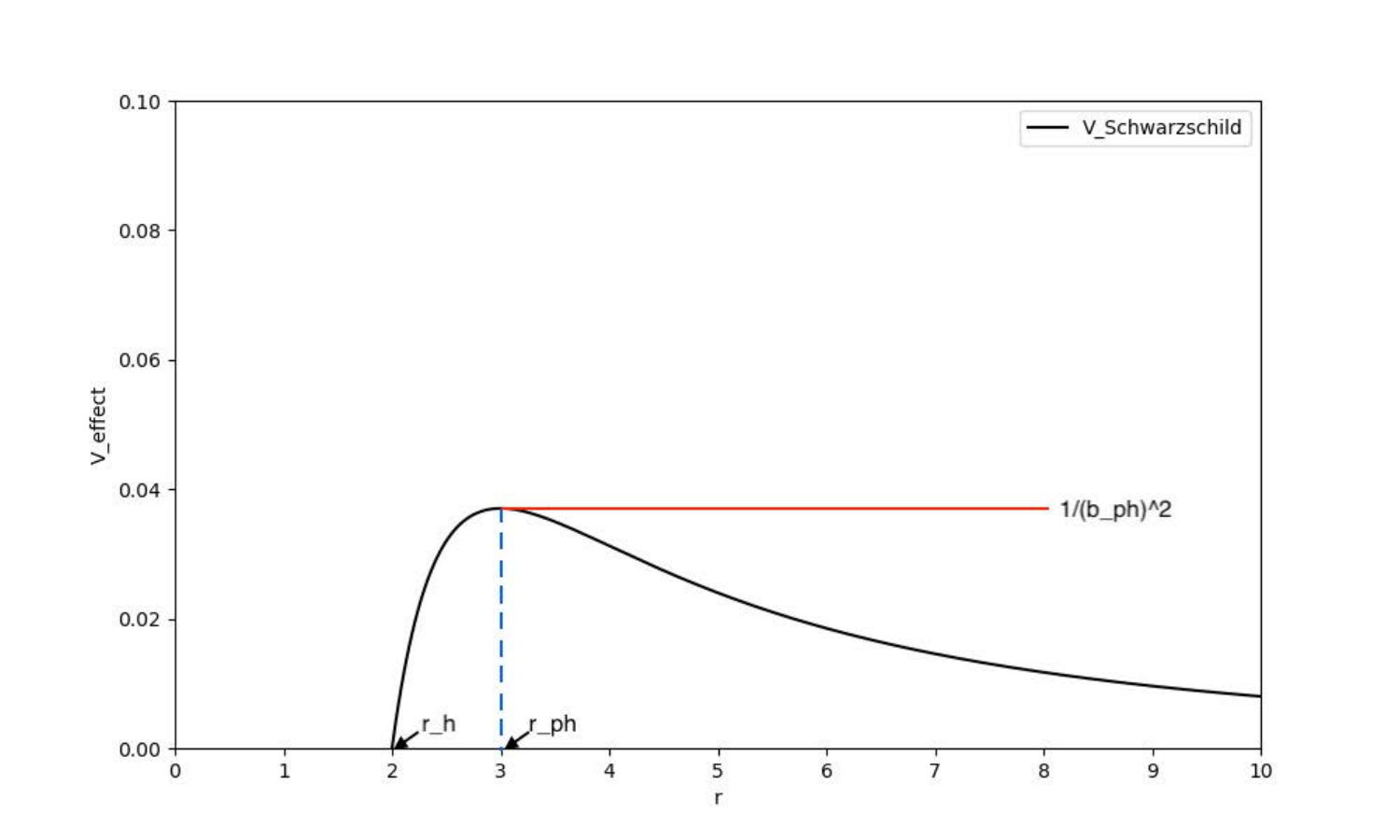}
\caption{The relationship between the effective potential $V_{\text{eff}}(r)$ and $r$ for the Schwarzschild black hole, as well as the relationship between $r_{\text{ph}}$ and $b_{\text{ph}}$.}
\label{fig:2}
\end{figure}

\begin{figure}[ht]
\centering
\includegraphics[width=0.5\textwidth]{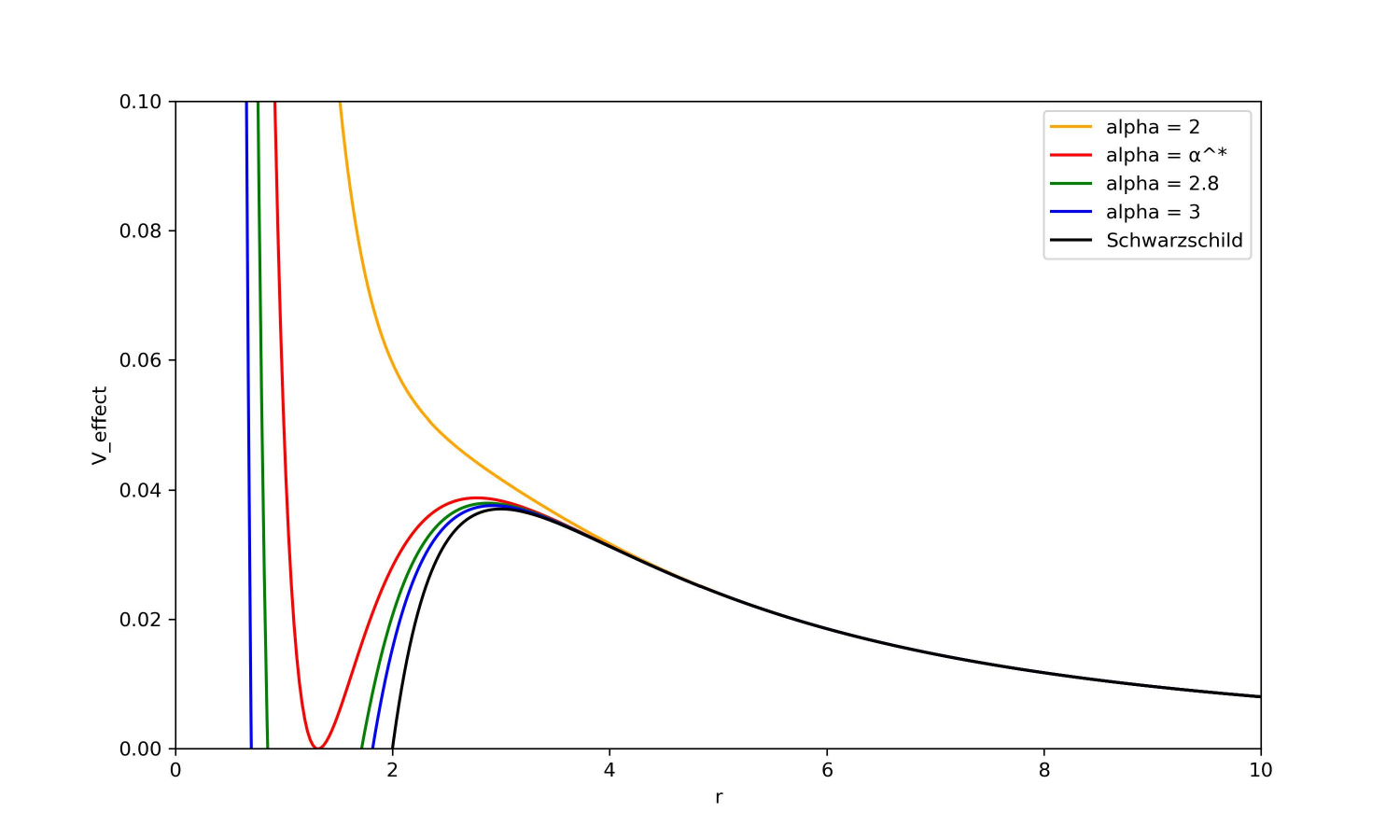}
\caption{The relationship curves of the effective potential \(V_{\text{eff}}(r)\) with respect to \(r\) are plotted for \(\alpha = 2\), \(\alpha = \alpha^*\), \(\alpha = 2.8\), \(\alpha = 3\), and the Schwarzschild case.}
\label{fig:3}
\end{figure}

For the spacetime with the line element \ref{1}, the geodesic equations can be derived from the Lagrangian

\begin{equation}
2\mathcal{L} = g_{ij} \frac{dx^i}{d\tau} \frac{dx^j}{d\tau},
\label{4}
\end{equation}

where $\tau$ is an affine parameter along the geodesic. For the spacetime metric (spherically symmetric four-dimensional spacetime), the Lagrangian is

\begin{equation}
\mathcal{L} = \frac{1}{2} \left[ -e^{\nu} \dot{t}^2 + e^{\lambda} \dot{r}^2 + r^2 \dot{\theta}^2 + (r^2 \sin^2 \theta) \dot{\phi}^2 \right].
\label{5}
\end{equation}
where the dot denotes a derivative with respect to $\tau$. The corresponding canonical momenta are
\begin{align}
p_t &= \frac{\partial L}{\partial \dot{t}} = -e^{\nu} \dot{t}, \\
p_{\phi} &= -\frac{\partial L}{\partial \dot{\phi}} = -r^2 \sin^2 \theta \dot{\phi}, \\
p_r &= -\frac{\partial L}{\partial \dot{r}} = -e^{\lambda} \dot{r}, \\
p_{\theta} &= -\frac{\partial L}{\partial \dot{\theta}} = -r^2 \dot{\theta}. 
\label{6}
\end{align}
Using the Legendre transformation $\mathcal{L} = p \dot{q} - \mathcal{H}$, where $p$ is the generalized momentum and $q$ is the generalized coordinate, the Hamiltonian is obtained as

\begin{equation}
\mathcal{H} = p_t \dot{t} + p_{\phi} \dot{\phi} + p_r \dot{r} + p_{\theta} \dot{\theta} - \mathcal{L} = \mathcal{L}. 
\label{7}
\end{equation}
Equation \ref{7} indicates that in the current problem, there is no "potential energy"; the energy comes solely from the "kinetic energy." The geodesic equations being derived correspond to a free particle system, and in general relativity, the world line of a free particle is its geodesic. Hence, the conservation of the Hamiltonian and Lagrangian is

\begin{equation}
\mathcal{H} = \mathcal{L} = \text{const}. 
\label{8}
\end{equation}
The Lagrangian of null geodesics must be zero. Further integrals of motion can be obtained from the following equations

\begin{equation}
\frac{dp_t}{d\tau} = \frac{\partial \mathcal{L}}{\partial t} = 0, \quad \frac{dp_{\phi}}{d\tau} = -\frac{\partial \mathcal{L}}{\partial \phi} = 0. 
\label{9}
\end{equation}
Thus,
\begin{equation}
p_t = -e^{\nu} \frac{dt}{d\tau} = \text{const.} \equiv E, 
\label{10}
\end{equation}
and
\begin{equation}
p_{\phi} = -r^2 \sin^2 \theta \frac{d\phi}{d\tau} = \text{const.} 
\label{11}
\end{equation}
where $E$ represents the energy of the system. Due to spherical symmetry, we can choose the geodesic to lie on a fixed plane, i.e., $\sin \theta = 1$, $\dot{\theta} = 0$, then Eq. \ref{11} becomes

\begin{equation}
p_{\phi} = -r^2 \frac{d\phi}{d\tau} = \text{const.} \equiv L.
\label{12}
\end{equation}
where $L$ denotes the angular momentum about the axis perpendicular to this plane. From Eqs. \ref{5}, \ref{10}, and \ref{12}, we obtain
\begin{equation}
\frac{E^2}{-e^{\nu}} + e^{\lambda} \dot{r}^2 + \frac{L^2}{r^2} = 2L = 0. 
\label{13}
\end{equation}
By considering $r$ as a function of $\phi$ (instead of $\tau$), i.e., $\frac{dr}{d\tau} = \frac{dr}{d\phi} \frac{d\phi}{d\tau}$, substituting into Eq. \ref{13}, and using $e^{\nu} \cdot e^{-\lambda} = 1$, we simplify to obtain

\begin{equation}
-E^2 + \frac{L^2}{r^4} \left( \frac{dr}{d\phi} \right)^2 + \frac{L^2}{r^2} e^{\nu} = 0. 
\label{14}
\end{equation}
Letting $u = r^{-1}$, we simplify to

\begin{equation}
\left( \frac{du}{d\phi} \right)^2 = \frac{1}{b^2} - e^{\nu(u)} u^2.
\label{15}
\end{equation}
where $b = \frac{L}{E}$ is the impact parameter. Equation \ref{15} is the null geodesic equation, determining the geometry of geodesics in the invariant plane. The effective potential is chosen as

\begin{equation}
V_{\text{eff}}(r) \equiv e^{\nu(u)} u^2, 
\label{16}
\end{equation}
Specifically, the circular null geodesic, referred to as the photon sphere, appears at the extremum of the effective potential \(V_{\text{eff}}(r)\). The existence of the photon sphere is conditioned by \cite{gan2021photon}

\begin{equation}
V_{\text{eff}}(r_{\text{ph}}) = \frac{1}{b_{\text{ph}}^2}, 
\label{17}
\end{equation}
and

\begin{equation}
V_{\text{eff}}'(r_{\text{ph}}) = 0.
\label{18}
\end{equation}
where $r_{\text{ph}}$ is the radius of the photon sphere, and $b_{\text{ph}}$ is the corresponding critical impact parameter. 
Their relationship is shown in Figure \ref{fig:2}, with the Schwarzschild case as an example, while the relationship between the effective potential \(V_{\text{eff}}(r)\) and \(r\) for different values of \(\alpha\) is depicted in Figure \ref{fig:3}.

From the Figure \ref{fig:2}, it can be seen that the event horizon radius of the Schwarzschild black hole is $r_h = 2$, and the radius of the photon sphere is $r_{\text{ph}} = 3$.

It is noteworthy that the local maximum of $V_{\text{eff}}$ ($\delta V_{\text{eff}} < 0$) and the local minimum point ($\delta V_{\text{eff}} > 0$) correspond to the radii of the unstable and stable photon spheres, respectively. The trajectory corresponding to the unstable photon sphere, when subjected to a slight perturbation, either falls into the black hole or escapes to infinity, observable by a distant observer. Conversely, the trajectory of the stable photon sphere does not exhibit such behavior. Therefore, the unstable photon sphere plays a crucial role in the image properties of the accretion disk observed by a distant observer \cite{gan2021photon}. The subsequent discussion will focus on the unstable photon orbits.

\section{\label{sec:level4}Fine Structure of the Photon Ring}
In this section, the method of backward ray tracing is employed to analyze the behavior of null geodesics near the hairy black hole with the Minkowski deformation. Additionally, the optical appearance of the black hole, when illuminated by an optically and geometrically thin accretion disk, is studied. The accretion disk is modeled as a thin plane with no absorption or scattering of photons, and background effects are not considered. The disk is assumed to be fixed on the equatorial plane (i.e., the plane perpendicular to the geodesics), with the observer positioned at the North Pole. Due to the strong gravitational field, null geodesics around the black hole may intersect with the accretion disk multiple times before either falling into the black hole or escaping to infinity, contributing differently to the total light intensity observed. Therefore, null geodesics are classified based on the number of intersections with the thin accretion disk, followed by the computation of the image of the hairy black hole with Minkowski deformation.
\subsection{\label{sec:level4.1}Classification of Null Geodesics}
The trajectory of null geodesics is closely related to the impact parameter \( b \) as shown by the geodesic equation \ref{15}. As described in the literature \cite{gralla2019black}, considering all null geodesics parallel to the direction of the North Pole, the range of the impact parameter can be classified into three categories based on the number of intersections between the null geodesics and the accretion disk, as given by the following expression
\begin{equation}
n = \frac{\phi}{2\pi}, 
\label{19}
\end{equation}
where \( \phi \) is the total change in the polar angle for the complete trajectory of the null geodesic in polar coordinates. For the case \( b < b_{\text{ph}} \), the total change in the polar angle outside the event horizon is given by \cite{gralla2019black, peng2021influence}

\begin{equation}
\phi = \int_0^{u_b} \frac{1}{\sqrt{\frac{1}{b^2} - e^{\nu(u)} u^2}} \, du, 
\label{20}
\end{equation}
where \( u_h \equiv 1/r_h \) and \( r_h \) is the radius of the event horizon. For \( b > b_{\text{ph}} \), the total change in the polar angle is given by \cite{yang2023shadow}

\begin{equation}
\phi = 2 \int_0^{u_{\text{max}}} \frac{1}{\sqrt{\frac{1}{b^2} - e^{\nu(u)} u^2}} \, du.
\label{21}
\end{equation}
where \( u_{\text{max}} \equiv 1/r_{\text{min}} \) and \( r_{\text{min}} \) is the minimum radial distance of the null geodesic to the black hole.

More specifically, if a null geodesic intersects the accretion disk plane only once, it corresponds to direct emission with \( 0 < n < \frac{3}{4} \); for the second type with \( \frac{3}{4} < n < \frac{5}{4} \), the null geodesic intersects the disk plane twice, corresponding to lensing ring emission; the third type corresponds to photon ring emission with \( n > \frac{5}{4} \), where the null geodesic intersects the disk plane at least three times. This classification is important because the number of intersections between the geodesic and the accretion disk is closely related to the observed intensity.

Figure \ref{4} shows the relationship between \( n \) and the impact parameter \( b \) for the Schwarzschild black hole. The three types are depicted in black, orange, and red, respectively.

\begin{figure}[ht]
\centering
\includegraphics[width=0.5\textwidth]{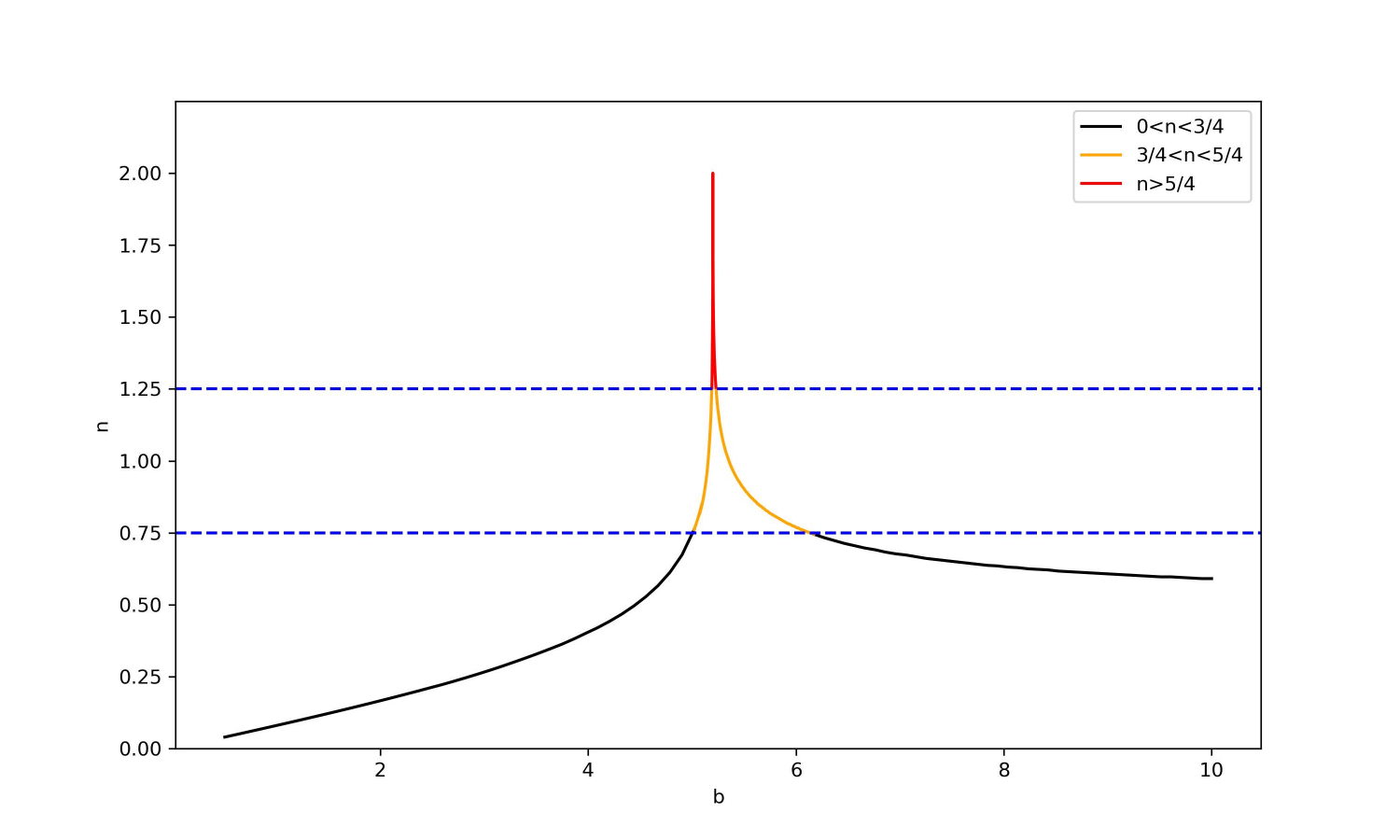}
\caption{The relationship between $n$ and the impact parameter $b$ for the Schwarzschild black hole. The intervals $0 < n < \frac{3}{4}$, $\frac{3}{4} < n < \frac{5}{4}$, and $n > \frac{5}{4}$ are depicted in black, orange, and red, respectively.}
\label{fig:4}
\end{figure}
From Figure \ref{4}, the intervals of the impact parameter for the Schwarzschild black hole can be classified as follows:

\begin{itemize}
\item \textbf{Direct Emission}: $0 < n < \frac{3}{4}$, corresponding to $b \in (0, b_2^-) \cup (b_2^+, +\infty)$.
\item \textbf{Lensing Ring Emission}: $\frac{3}{4} < n < \frac{5}{4}$, corresponding to $b \in (b_2^-, b_3^-) \cup (b_3^+, b_2^+)$.
\item \textbf{Photon Ring Emission}: $n > \frac{5}{4}$, corresponding to $b \in (b_3^-, b_3^+)$.
\end{itemize}

Using the null geodesic equation \ref{15}, the relationship between $n$ and the impact parameter $b$ is shown in Figure 4. Here, $\alpha$ is chosen as $\alpha^*$, 2.8, 3, and the Schwarzschild solution, with the four cases compared.

\begin{figure}[ht]
\centering
\includegraphics[width=0.5\textwidth]{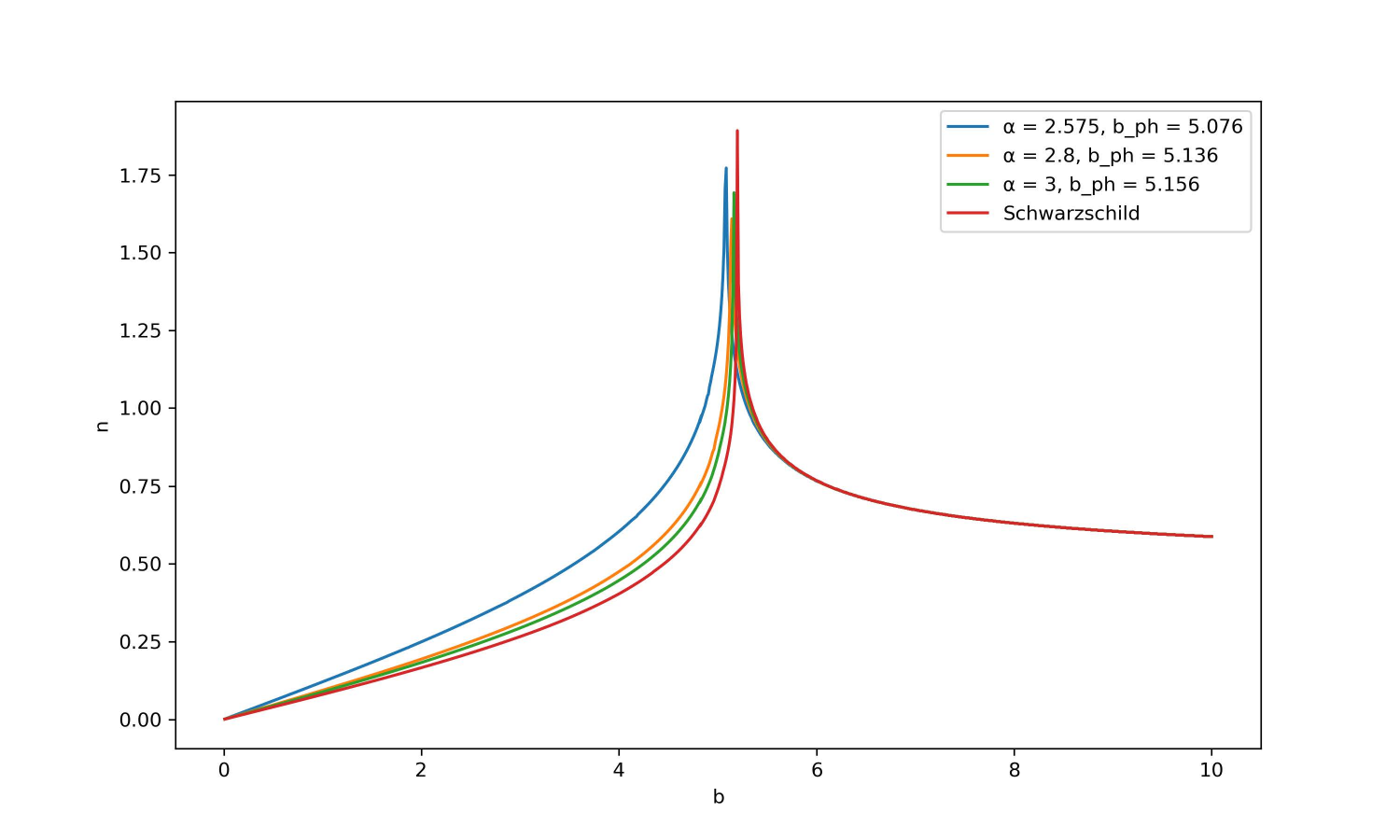}
\caption{The relationship between $n$ and the impact parameter $b$ for different values of $\alpha$: $\alpha^*$ (blue), $2.8$ (orange), $3$ (green), and the Schwarzschild solution (red).}
\label{fig:5}
\end{figure}

From Figure \ref{fig:5}, we observe that in all cases, as the impact parameter $b$ increases, the bending of the null geodesics initially increases, reaches infinity at $b = b_{\text{ph}}$, and then gradually decreases. As $\alpha$ increases, $b_{\text{ph}}$ also increases and eventually approaches the value of $3\sqrt{3}$, which corresponds to the Schwarzschild black hole. The Schwarzschild solution serves as a benchmark, and in the limit $\alpha \rightarrow +\infty$, the $n$-$b$ relationship curve returns to the Schwarzschild case. This behavior is expected. In fact, we can see that when $\alpha = 3$, the $n$-$b$ relationship curve is already very close to the Schwarzschild case.

It is noteworthy that as $\alpha$ increases from $\alpha^*$, the influence of $\alpha$ on various physical properties of the photon ring gradually decreases, which will also be evident in subsequent sections. Additionally, with the increase of the parameter $\alpha$, the range of the impact parameter $b$ corresponding to the lensing ring and the photon ring becomes narrower.

\begin{figure}[ht]
\centering
\includegraphics[width=0.5\textwidth]{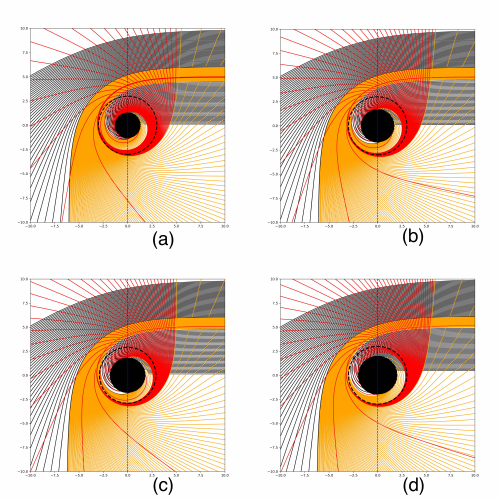}
\caption{Null geodesics around the black hole for different values of $\alpha$: $\alpha^*$ (a), $2.8$ (b), $3$ (c), and the Schwarzschild solution (d). The black disk corresponds to the region within the event horizon, the circular dashed black line represents the radius of the unstable photon sphere, and the vertical blue dashed line indicates the equatorial plane where the accretion disk is located. The observer is located at infinity to the right (North Pole). The curves represent null geodesics bending around the black hole. The color coding corresponds to the number of intersections with the accretion disk: black for one intersection, orange for two intersections, and red for three or more intersections.}
\label{fig:6}
\end{figure}

Below, using the null geodesic equation \ref{15} and based on the corresponding physical assumptions, we can plot the following null geodesic diagram Figue \ref{fig:6}.

From Figure \ref{fig:6}, we can deduce the following:
1. The physical significance of the impact parameter \( b \) in the figure is the coordinate value corresponding to the null geodesic projected to the right (North Pole) at infinity. This indicates the radial distance from the center of the black hole to the point where the observer receives the light ray at infinity.
2. In all cases, as the impact parameter \( b \) increases, the bending of the corresponding null geodesic (number of intersections with the accretion disk) initially increases, reaches infinity at \( b = b_{\text{ph}} \), and then gradually decreases. The color transition follows the sequence black → orange → red → orange → black. Thus, the analysis in Figure \ref{fig:4} can be summarized as follows: As \( \alpha \) increases, the radial distance from the center of the black hole to the observer's position at infinity for the most bent geodesic (corresponding to the photon sphere) increases gradually to the value $3\sqrt{3}$, which is the case for the Schwarzschild black hole.
3. As the parameter \( \alpha \) increases from the extreme case \( \alpha^* \), the event horizon radius \( r_h \), photon sphere radius \( r_{\text{ph}} \), and critical impact parameter \( b_{\text{ph}} \) all change (gradually increase).

By further precise calculations, we determine the boundaries of the impact parameter \( b \) corresponding to different numbers of intersections with the accretion disk. The results are listed in Table \ref{tab:1}, which also shows the event horizon radius \( r_h \), photon sphere radius \( r_{\text{ph}} \), critical impact parameter \( b_{\text{ph}} \), and the innermost stable circular orbit \( r_{\text{isco}} \) (which will be used later). The innermost stable circular orbit \( r_{\text{isco}} \) can be calculated via Eq. \ref{22}\cite{wang2023rings}

\begin{equation}
r_{\text{isco}} = \frac{3e^{\nu(r_{\text{isco}})} \left( e^{\nu(r_{\text{isco}})} \right)'}{2 \left( \left( e^{\nu(r_{\text{isco}})} \right)' \right)^2 - e^{\nu(r_{\text{isco}})} \left( e^{\nu(r_{\text{isco}})} \right)''}.
\label{22}
\end{equation}
where primes denote derivatives with respect to the radial coordinate $r$.

\begin{table}[ht]
\centering
\begin{tabular}{ccccccccc}
\hline
\hline
$\alpha$ & $b_2^-$ & $b_2^+$ & $b_3^-$ & $b_3^+$ & $r_h$ & $r_{\text{ph}}$ & $b_{\text{ph}}$ & $r_{\text{isco}}$ \\
\hline
$\alpha^*$ & 4.550 & 6.136 & 5.012 & 5.140 & 1.331 & 2.777 & 5.080 & 5.978 \\
2.8       & 4.840 & 6.145 & 5.112 & 5.183 & 1.716 & 2.878 & 5.135 & 5.992 \\
3         & 4.920 & 6.160 & 5.144 & 5.199 & 1.817 & 2.904 & 5.160 & 5.997 \\
Schwarzschild & 5.020 & 6.170 & 5.190 & 5.230 & 2.000 & 3.000 & $3\sqrt{3}$ & 6.000 \\
\hline
\hline
\end{tabular}
\caption{Boundary values of the impact parameter $b$ for different $\alpha$ values ($\alpha^*$, 2.8, 3, and the Schwarzschild solution), along with the event horizon radius $r_h$, photon sphere radius $r_{\text{ph}}$, critical impact parameter $b_{\text{ph}}$, and the innermost stable circular orbit $r_{\text{isco}}$.}
\label{tab:1}
\end{table}

\subsection{\label{sec:level4.2}Re-Constraining the Parameter \(\alpha\)}

We can constrain the parameter \(\alpha\) in the metric of the hairy black hole based on the observed shadow and photon ring radii of the black holes M87* and Sagittarius A* (SgrA*). According to the Event Horizon Telescope (EHT) collaboration's studies on the structure of the photon rings of these black holes \cite{collaboration2019first, akiyama2022first}, the angular diameter of the shadow of M87* is \(\theta_{\text{M87}^*} = (42 \pm 3) \, \mu \text{as}\), with a distance from Earth of \(D_{\text{M87}^*} = 16.8^{+0.8}_{-0.7} \, \text{Mpc}\) and a mass of \(M_{\text{M87}^*} = (6.5 \pm 0.9) \times 10^9 M_{\odot}\). For SgrA*, the angular diameter of the shadow is \(\theta_{\text{SgrA}^*} = (48.7 \pm 7) \, \mu \text{as}\), with a distance from Earth of \(D_{\text{SgrA}^*} = 8277 \pm 33 \, \text{pc}\) and a mass of \(M_{\text{SgrA}^*} = (4.3 \pm 0.013) \times 10^6 M_{\odot}\). Therefore, the diameters of the shadow images of M87* and SgrA* are \(d_{\text{sh}}^{\text{M87}^*} = 11 \pm 1.5\) and \(d_{\text{sh}}^{\text{SgrA}^*} = 9.5 \pm 1.4\), respectively \cite{luo2024shadows}.

For a static observer at position \(r_0\), the shadow radius of the black hole can be expressed as \cite{psaltis2020gravitational}

\begin{equation}
r_{\text{sh}} = \frac{r_{\text{ph}}}{\sqrt{-g_{tt}(r_{\text{ph}})}}.
\label{23}
\end{equation}
Using Eq. \eqref{2}, we obtain \(r_{\text{sh}} = \frac{r_{\text{ph}}}{\sqrt{e^{\nu(r_{\text{ph}})}}}\). Due to symmetry, the diameter of the black hole shadow is \(d_{\text{sh}} = 2r_{\text{sh}}\). 

\begin{figure}
\centering
\includegraphics[width=0.5\textwidth]{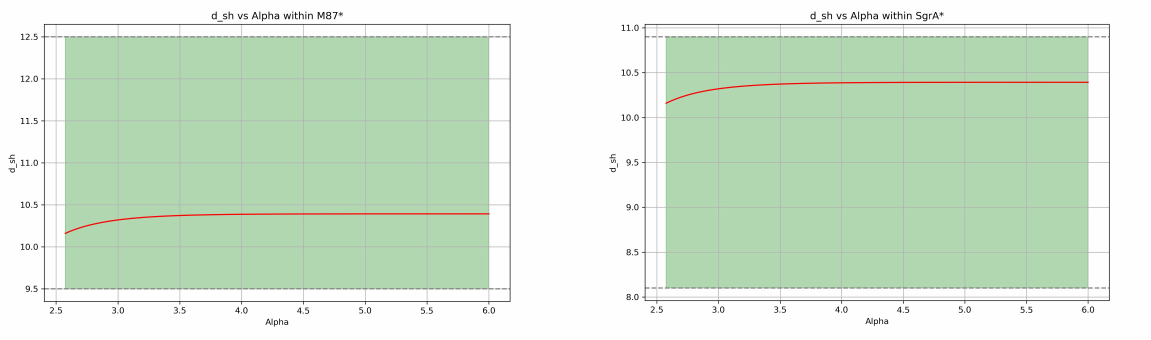}
\caption{Constraints on the parameter \(\alpha\) based on the observed diameters of the shadows of M87*(left) and SgrA*(right) from the EHT collaboration. The red curve represents the trend of the black hole shadow diameter with varying \(\alpha\), and the green shaded region indicates the possible range of the observed black hole shadow diameter.}
\label{fig:7}
\end{figure}

In Figure \ref{fig:7}, we illustrate the constraints on the parameter \(\alpha\) based on the observed diameters of the shadows of M87* and SgrA* as obtained by the EHT collaboration.

From Figure \ref{fig:7}, it is evident that for both M87* and SgrA* black holes, the parameter \(\alpha\) can range from \(\alpha^* (\approx 2.575)\) to positive infinity (approaching the Schwarzschild black hole scenario).

\subsection{\label{sec:level4.3}Observed Intensity and Optical Appearance}

Next, we simulate the images of the photon ring and the shadow of the black hole as seen by an observer at infinity for different values of the parameter \(\alpha\). As previously mentioned, we consider the accretion disk as the sole light source. For those null geodesics intersecting the accretion disk, the light emitted from the disk will travel along these geodesics to the observer. Thus, each intersection with the accretion disk extracts energy, and the more intersections, the more energy is extracted, resulting in a higher brightness, which directly affects the distribution of the observed intensity.

Assuming that the accretion disk emits isotropically in the rest frame of the static black hole with an emission frequency \(\nu_e\), the observed specific intensity at a frequency \(\nu_0\) can be obtained using Liouville's theorem \cite{jaroszynski1997optics, gan2021photon, bromley1997line}:

\begin{equation}
I_o (r, \nu_0) = g^3 I_e (r, \nu_e).
\label{24}
\end{equation}
where \(g = \frac{\nu_0}{\nu_e} = \sqrt{e^{\nu(r)}}\) is the redshift factor, \(\nu_0\) and \(\nu_e\) are the observed and emitted frequencies, respectively, and \(I_o (r, \nu_0)\) and \(I_e (r, \nu_e)\) are the observed and emitted specific intensities at a given radius. Integrating over all observed frequencies, we obtain the total observed intensity:

\begin{equation}
I_{\text{obs}} (r) = \int I_o (r, \nu_0) d\nu_0 = \int g^4 I_e (r, \nu_e) d\nu_e = (e^{\nu(r)})^2 I_{\text{em}} (r).
\label{25}
\end{equation}
where \(I_{\text{em}} (r) = \int I_e (r, \nu_e) d\nu_e\) is the total emitted intensity. However, as mentioned earlier, because null geodesics may intersect the accretion disk multiple times, the total observed intensity is the sum of the intensities at each intersection \cite{gralla2019black}

\begin{equation}
I_{\text{obs}} (b) = \sum_m \left. (e^{\nu(r)})^2 I_{\text{em}} (r) \right|_{r = r_m (b)}.
\label{26}
\end{equation}
where \(r_m (b)\) is the transfer function, representing the radial coordinate of the \(m\)-th intersection (where \(m = 1, 2, 3, \ldots\)) of the null geodesic with impact parameter \(b\) with the accretion disk. According to the backward ray tracing method, the \(m\)-th intersection is named based on the sequential order in which the null geodesic intersects the accretion disk from the observer’s perspective. The slope \(dr_m / db\) for a fixed m of the transfer function describes the demagnification or magnification factor, with larger \(m\) corresponding to strong demagnification or magnification.

In fact, from the null geodesic plot of the black hole in Figure \ref{fig:6}, we see that when \(m \geq 3\) (corresponding to the red curves), the radial distance of the third intersection varies significantly with \(b\). This means that in the vicinity of \(b_r\) (near \(b_r^+\) and \(b_r^-\)), the energy extracted by the null geodesics from the accretion disk changes dramatically, corresponding to very strong demagnification. It is noteworthy that Equation \ref{26} involves a variable substitution (from \(r\) to \(b\)) compared to Equation \ref{25}, with the purpose of simulating the image of the photon ring and its shadow as seen by an observer at infinity. For such an observer, the impact parameter \(b\) represents the radial distance of the observed bright ring from the center of the black hole.

\begin{figure}[ht]
\centering
\includegraphics[width=0.5\textwidth]{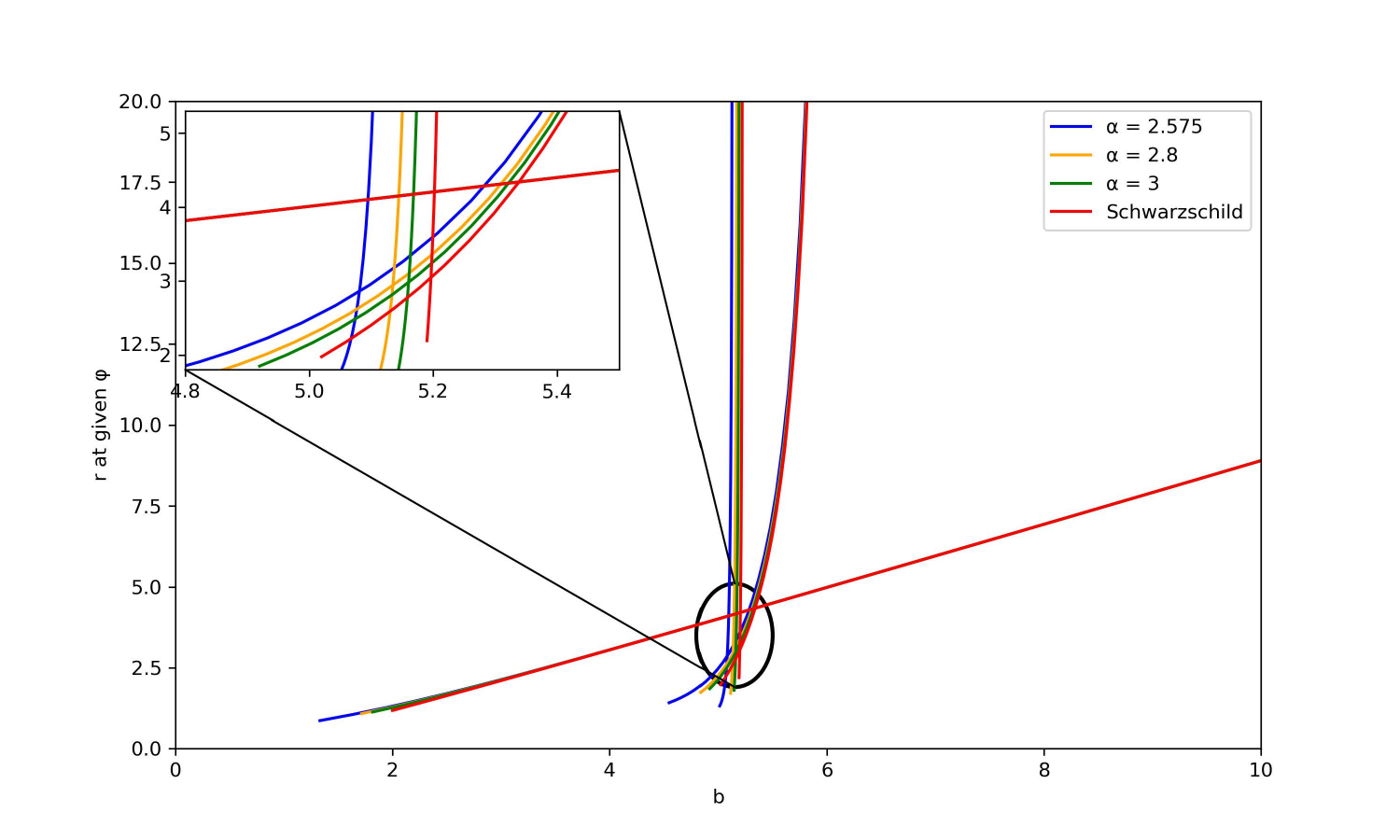}
\caption{The first three transfer functions for different values of \(\alpha\): \(\alpha^*\) (blue), 2.8 (orange), 3 (green), and the Schwarzschild solution (red).}
\label{fig:8}
\end{figure}

By definition, \(m \leq 1\), \(m = 2\), and \(m \geq 3\) correspond to direct emission, lensing ring emission, and photon ring emission, respectively. For \(m > 3\), due to strong demagnification, their contribution to the total luminosity can be ignored \cite{gralla2019black}. As long as the local emission does not differ significantly from the rest of the accretion disk, the main contribution to brightness comes from the lensing ring \cite{gralla2019black}. The quantitative relationships of the first three transfer functions can be described as follows \cite{yang2023shadow}

\begin{equation}
r_1 (b) = \frac{1}{u(\pi/2, b)}, \quad b \in (b_1^-, +\infty), 
\label{27}
\end{equation}

\begin{equation}
r_2 (b) = \frac{1}{u(3\pi/2, b)}, \quad b \in (b_2^-, b_2^+), 
\label{28}
\end{equation}

\begin{equation}
r_3 (b) = \frac{1}{u(5\pi/2, b)}, \quad b \in (b_3^-, b_3^+). 
\label{29}
\end{equation}
where \(u(\phi, b)\) is the solution of the geodesic equation \ref{15}. Figure \ref{fig:8} shows the first three transfer functions for different values of \(\alpha\), specifically \(\alpha^*\), 2.8, 3, and the Schwarzschild solution.

From Figure \ref{fig:8}, we observe that as \(\alpha\) increases, the regions of the three transfer functions gradually narrow and approach the Schwarzschild case. In all cases, the slope of the transfer functions becomes steeper with increasing \(m\), corresponding to the previously mentioned strong demagnification for larger \(m\). It is easy to see that the slopes corresponding to all \(r_1(b)\) are nearly 1, indicating that the radial coordinate of the first intersection of the null geodesic with the accretion disk varies almost linearly with the impact parameter.

To further study the optical appearance of the hairy Minkowski-deformed black hole, we need to specify the emission intensity \(I_{\text{em}}(r)\) of the accretion disk, which represents the function of the light intensity emitted from the accretion disk as a function of distance from the black hole. We consider the following three specific toy models of emission intensity \cite{li2021shadows}:

\begin{itemize}
\item \textbf{First emission model, with an inverse-square law}: 
\begin{equation}
I_{\text{em}}(r) = 
\begin{cases} 
	I_0 \left[\frac{1}{r - (r_{\text{isco}} - 1)}\right]^2, & r > r_{\text{isco}} \\
	0, & r \leq r_{\text{isco}}
\end{cases} ,
\label{30}
\end{equation}
\item \textbf{Second emission model, with an inverse-cube law}: 
\begin{equation}
I_{\text{em}}(r) = 
\begin{cases} 
	I_0 \left[\frac{1}{r - (r_{\text{ph}} - 1)}\right]^3, & r > r_{\text{ph}} \\
	0, & r \leq r_{\text{ph}}
\end{cases} ,
\label{31}
\end{equation}
\item \textbf{Third emission model, with a more gradual attenuation}: 
\begin{equation}
I_{\text{em}}(r) = 
\begin{cases} 
 I_0 \frac{\left(\frac{\pi}{2} - \tan^{-1}[r - (r_{\text{isco}} - 1)]\right)}{\left(\frac{\pi}{2} - \tan^{-1}[r_h - (r_{\text{isco}} - 1)]\right)}, & r > r_h \\
 0, & r \leq r_h
\end{cases} .
\label{32}
\end{equation}
\end{itemize}
where \(r_h\) is the event horizon radius, \(r_{\text{ph}}\) is the photon sphere radius, \(b_{\text{ph}}\) is the critical impact parameter, and \(r_{\text{isco}}\) is the innermost stable circular orbit, with their respective values given in Table \ref{tab:1}.

\begin{figure*}
\includegraphics[width=1\textwidth]{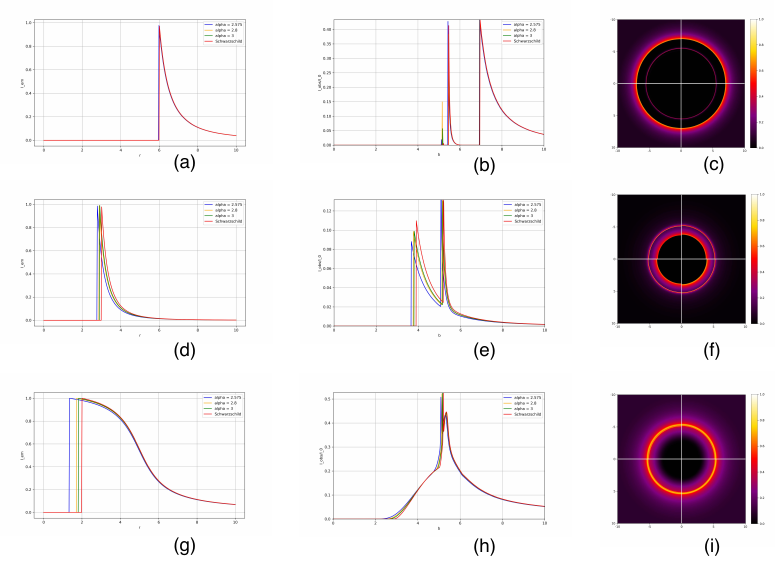}
\caption{The total observed intensity function curves and optical appearances of the shadows and rings of the Minkowski-deformed black hole for different values of \(\alpha\): \(\alpha^*\), 2.8, 3, and the Schwarzschild solution. Panels (a), (b), and (c) correspond to the first emission model, panels (d), (e), and (f) correspond to the second emission model, and panels (g), (h), and (i) correspond to the third emission model. For clarity, panels (c), (f), and (i) are composed of images for \(\alpha = \alpha^*\) (top left), 2.8 (top right), 3 (bottom left), and the Schwarzschild solution (bottom right). The color bars have been normalized.}
\label{fig:9}
\end{figure*}

Using these emission intensity functions and transfer functions, we substitute them into equation \ref{26} and plot Figure \ref{fig:9}, which visually shows the shadows and rings of the Minkowski-deformed black hole for \(\alpha\) values of \(\alpha^*\), 2.8, 3, and the Schwarzschild solution. First, we plot the total intensity curves as functions of the impact parameter for different \(\alpha\) values, and then, due to symmetry, we project them onto a two-dimensional plane and apply color mapping to clearly present the optical appearance of the accretion disk.

From Figure \ref{fig:9}, we observe the following:
1. In all three toy models, as \(\alpha\) increases, the peak of the total observed intensity function curve shifts toward larger \(b\) values, indicating that larger photon ring radii are observed for larger \(\alpha\). This trend suggests that the photon ring images of Minkowski-deformed hairy black holes are not degenerate with respect to \(\alpha\), providing a means to test the no-hair theorem. Different photon ring radii correspond to different values of \(\alpha\), allowing the optical appearance of the photon ring to reveal information about this type of black hole.
2. In the observed intensity function curves for the first and second models, two distinct peaks are visible, resulting in two bright rings in the two-dimensional projection of the total observed intensity. The third model also shows two bright rings, although they are more closely spaced and less distinguishable.
3. In the first and third models, the optical appearance of the photon ring for different \(\alpha\) values is not significantly different. However, the second model better reflects the properties of this type of black hole. This is because the emission function of the second model is determined by the photon sphere radius \(r_{\text{ph}}\), making the optical appearance of the photon ring more sensitive to changes in \(\alpha\).

\section{\label{sec:level5}Conclusion and Discussion}

The release of black hole images by the Event Horizon Telescope (EHT) collaboration has opened a new window for studying null geodesics in strong gravitational fields and provided new insights into the possible spacetime structure around black holes. This paper investigates the optical appearance of the Minkowski-deformed hairy black hole, constructed using the gravitational decoupling (GD) method, under the illumination of three toy models of emission intensity. The influence of the hairy parameter \(\alpha\) on the optical appearance is discussed.

In the study, we first analyzed the effect of \(\alpha\) on the black hole event horizon. We found that when \(\alpha = \alpha^*\), the black hole has an event horizon; when \(\alpha\) is smaller, the black hole does not have an event horizon, resulting in a naked singularity; when \(\alpha > \alpha^*\), the horizon function has two zeros, corresponding to the Cauchy horizon and the event horizon. Based on the cosmic censorship hypothesis, which prohibits the appearance of naked singularities, we investigated the optical appearance of the photon ring of the Minkowski-deformed black hole within the range \(\alpha^* \leq \alpha < +\infty\). Using the effective potential, we determined the radius of the unstable photon sphere and, with the backward ray-tracing algorithm, studied the null geodesics around this type of black hole. We found that the distribution and classification of these geodesics are influenced by \(\alpha\). As \(\alpha\) increases, the range of the impact parameter \(b\) corresponding to the lensing ring and the photon ring narrows, affecting the image of the photon ring. In the study by Qingyu Gan et al. \cite{gan2021photon}, it was found that hairy black holes possess at least one unstable photon sphere, and in certain parameter ranges, a stable photon sphere may also appear. According to Figure \ref{fig:3}, this paper indeed identifies the presence of an unstable photon sphere in the Minkowski-deformed hairy black hole but does not observe a stable photon sphere outside the event horizon. As mentioned in the introduction, the stability of various black holes, such as colored black holes, Bartnik-McKinnon black holes, and Yang-Mills black holes, has been extensively studied. These studies reveal that not all black holes are inherently stable\cite{straumann1990instability, bizon1991n, lavrelashvili1995remark, bizon1991stability}. According to \cite{cardoso2014light}, this might be due to the stability of the spacetime considered in this paper.

We then investigated the optical appearance of the Minkowski-deformed black hole illuminated by an optically and geometrically thin accretion disk. Assuming the accretion disk radiates according to the three toy models \ref{30}, \ref{31}, and \ref{32}, we found that the total observed intensity always has two distinct peaks for the first and second models, corresponding to two bright rings in the two-dimensional image. For all models, larger \(\alpha\) values result in larger photon ring radii, eventually converging to the Schwarzschild case. However, compared to the other two models, the optical appearance of the photon ring of the Minkowski-deformed black hole is more sensitive in the second emission model, making it useful for future studies.

In summary, this paper uses the photon ring and shadow images of the Minkowski-deformed hairy black hole to explore the influence of the hairy parameter \(\alpha\). The results show that the photon ring and images can be used to determine the properties of this type of spacetime. In theory, photon rings can distinguish different spacetime metrics of Minkowski-deformed regular hairy black holes, providing a potential method for testing the no-hair theorem through future observations. However, it should be noted that photon rings may not distinguish all spacetime properties. As Yuan Meng et al. \cite{meng2023images} found, the competition between two hairy parameters (\(\alpha\) and \(lo\)) for a hairy Schwarzschild black hole with a single photon sphere can potentially lead to degenerate photon ring images. In the study by Kostas Glampedakis et al. \cite{glampedakis2023black}, it was shown that dark objects with strongly non-Kerr multipole structures can produce Kerr-like shadows with characteristic quasi-circular shapes, resulting in significant degeneracy when considering multipole moments beyond the quadrupole. These findings suggest that in certain spacetimes, black hole shadows or photon rings may not reliably test the no-hair theorem. Understanding when photon rings reliably represent spacetime and when they do not involves more fundamental and intriguing black hole physics.

We hope to inspire insights and meaningful efforts in the study of black hole shadows and the no-hair theorem. In the future, we will seek to understand the behavior of photon rings in black holes with charge or axial symmetry and discuss the possibility of photon rings serving as holographic horizons within the framework of holographic duality for astrophysical black holes.

\section{acknowledgements}
We acknowledge the anonymous referee for a constructive report that has significantly improved this paper. This work was supported by the Special Natural Science Fund of Guizhou University (Grant No.X2022133), the National Natural Science Foundation of China (Grant No. 12365008) and the Guizhou Provincial Basic Research Program (Natural Science) (Grant No. QianKeHeJiChu-ZK[2024]YiBan027) . 


\end{document}